\documentclass[a4paper]{jpconf}
\usepackage{graphicx}
\usepackage{amsmath}
\usepackage{amssymb}
\usepackage{slashed}
\usepackage[colorlinks,citecolor=blue,linkcolor=red,urlcolor=black]{hyperref}
\begin{document}
\title[Magnetic FESR]{Magnetic effects of QCD parameters from finite energy sum rules}

\author{Cristi\'an Villavicencio}

\address{Centro de Ciencias Exactas and Departamento de Ciencias B\'asicas, Universidad del Bío-Bío, Avda. Andrés Bello 720,
Casilla 447, 3800708, Chillán, Chile.
}

\ead{cvillavicencio@ubiobio.cl}

\author{C. A. Dominguez} 
\address{Centre for Theoretical \& Mathematical Physics, and Department of Physics, University of Cape Town,
Rondebosch 7700, South Africa.}

\author{M. Loewe}
\address{Instituto de Fisica, Pontificia Universidad Catolica de Chile,
Casilla 306, Santiago, Chile;}
\address{Centro Científico Tecnol\'ogico de Valparaíso-CCTVAL,
Universidad T\'ecnica Federico Santa Mar\'ia, Casilla 110-V, Valpara\'iso, Chile;
Centre for Theoretical \& Mathematical Physics, and Department of Physics, University of Cape Town,
Rondebosch 7700, South Africa.}

\begin{abstract}
One of the advantages of the finite energy sum rules is the fact that every operator in the operator product expansion series can be selected individually by the use of an appropriate kernel function which removes other operator poles. 
This characteristic is maintained by QCD systems in the presence of external homogeneous magnetic field, providing interesting information about the magnetic evolution of QCD and hadronic parameters. 
In this work finite energy sum rules are applied on  QCD in the light quark sector, combining axial and pseudoscalar channels in the presence of an external homogeneous magnetic field, obtaining the magnetic evolution of the light quark masses, pion mass, the pion decay constant, the gluon condensate and the continuum hadronic threshold. 

\end{abstract}

\section{Introduction}
The increasing interest in the study of magnetic field effects in quantum chromodynamics (QCD), mainly in the search of the chiral magnetic effect in relativistic heavy ion collision experiments, as well as the interior of magnetars, has produced a variety of interesting and different techniques in order to obtain the modification of particle parameter due to the interaction with the medium. 
The main difficulty arises from trying to cover all the possible range of the magnetic field strength. A typical approximation is the use of the lowest Landau level, which  is the main tool to explore  QCD under highly magnetized conditions.
However, it commonly exceeds the physical possible values that collider experiments or magnetars can reach. Another approximation is to expand in power series of the magnetic field, with the range of magnetic field values restricted by the scale of the model. 
In this work, the finite energy sum rules (FESR) will be employed in the presence of an external magnetic field in order to obtain the magnetic evolution for hadronic and QCD parameters. 
More references and details an be found in \cite{Dominguez:2018njv}, which is the article on which this text is mainly based.

\begin{figure}
\begin{minipage}{.55\textwidth}
\includegraphics[scale=.49]{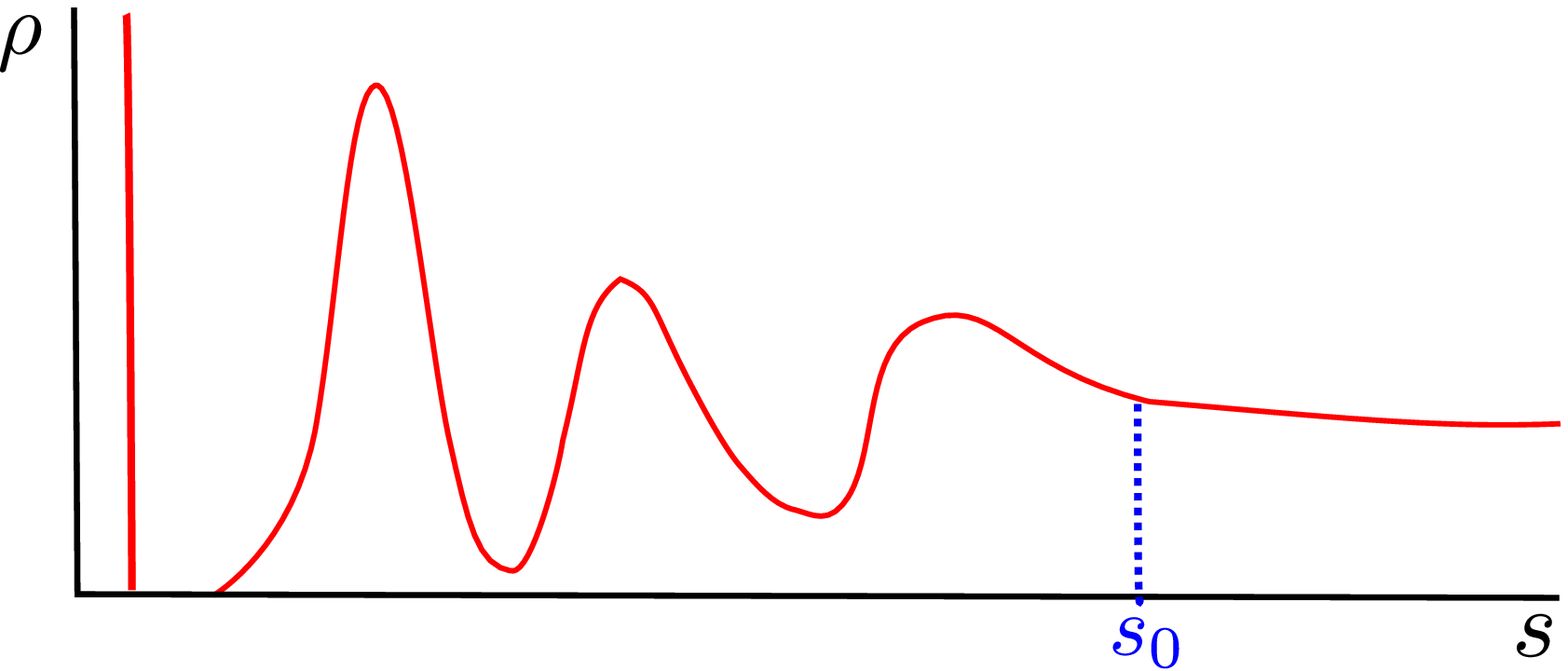}
\caption{Schematic representation of hadronic resonances. $s_0$ denotes the limit where heavier and broader resonances start to overlap.}
\label{resonances}
\end{minipage}
\hspace{.03\textwidth}
\begin{minipage}{.41\textwidth}
\includegraphics[scale=.6]{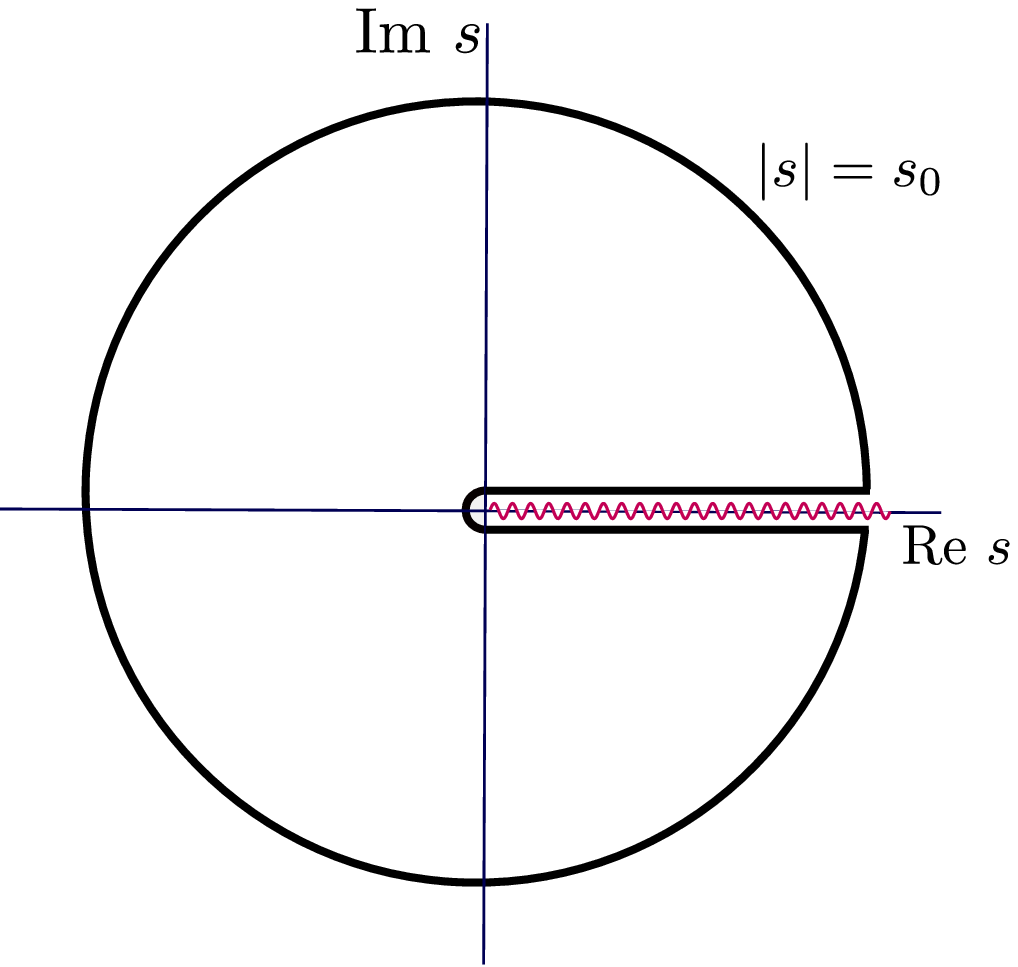}
\caption{ The \emph{pac-man} contour. Hadronic resonances lies over the real axis, while QCD lives on the circle}
\label{pacman}
\end{minipage}
\end{figure}

The FESR relates the hadronic with the quarks sector based on analytic properties in the complex squared-energy $s$-plane.  
Let us consider the correlation of two currents in Fourier space
\begin{equation}
\Pi(p^2)=\int d^4x\, e^{ip\cdot x}\langle 0| TJ(x)J^\dag(0)|0\rangle.
\end{equation}
In the hadronic sector, the spectral density function can be obtained as $\rho(s)=\frac{1}{\pi}\mathrm{Im}\,\Pi(s+i\epsilon)$. 
Figure \ref{resonances} shows a scheme of the spectral function as a function of the squared energy, where the stable particle state can be observed as a delta function and where a set of resonances appear at high energy. 
The parameter $s_0$ denotes the threshold where broader resonances start to overlap. 
This parameter is known as the \emph{continuum hadronic  threshold}.

 The only singularities of current correlators  lie on the right-hand complex plane. 
They are in the form of poles on the real s-axis (stable hadrons), or on the second Riemann sheet (hadronic resonances).

Next, Cauchy's theorem is invoked in the s-plane, considering the contour shown on the right side of Fig. \ref{pacman}, leading to finite energy QCD sum rules (FESR).
\begin{equation}
\frac{1}{\pi} \int_{0}^{s_0} ds\, K(s)\, {\mbox{Im}}\, \Pi(s)|_\text{\tiny{Had}} = 
  \frac{-1}{2 \pi i} \oint_\text{\tiny{C($s_0$)}} ds\,K(s)\,\Pi(s)|_\text{\tiny{QCD}},  \label{FESR}
\end{equation}
where $K(s)$ is an analytic kernel. 
We use here $K(s)=s^{N-1}$, with $N$ an integer number, with the advantage that it cuts the operator product expansion (OPE) of current correlators in QCD.
This operator series parameterize the non perturbative sector in terms of vacuum expectation values of different operators of dimension $2n$,
\begin{equation}
    \Pi(s)|_\text{\tiny QCD} = \sum_{n=0}C_{2n} (s,\mu^2)\frac{\langle \hat{\mathcal{O}}_{2n} (\mu^2) \rangle }{(-s)^{n}} , \label{OPE}
\end{equation}
where $\mu^2$ is a renormalization scale and $C_{2n}$ are the Wilson coefficients.
The $n=0$ case corresponds to perturbative QCD which is usually a logarithmic contribution. 
The lowest dimension operators are ${\cal O}_4$, which corresponds the quark condensate (multiplied by the quark mass) and the gluon condensate.  
From here is easy to see that, if Wilson coefficients are constant, there will be only one specific operator contribution for each kernel used. 

While the Wilson coefficients in the OPE, Eq.~(\ref{OPE}) can be computed in PQCD, the values of the vacuum condensates cannot be obtained analytically from first principles. 
These condensates can be determined from the QCD sum rules themselves, in terms of some experimental information input, by lattice QCD (LQCD) simulations or information provided from other models.

We consider three current correlators in the presence of an external magnetic field: the light-quark axial-vector current correlator, pseudoscalar correlator and axial-pseudoscalar correlator.

\section{Vacuum FESR}
\label{sec:vacCorr}
The axial-vector current correlator is defined as
\begin{equation}
\Pi_{\mu\nu}^{A} (p) \; = \; i  \int  d^{4} x \, e^{i p\cdot x} \,
\langle 0|T [A_{\mu}(x)  A_{\nu}^{\dagger}(0)]|0 \rangle 
\;=\; p_\mu p_\nu \, \Pi_0(p^2) +g_{\mu\nu}\, \Pi_1(p^2) 
\end{equation}
where $A_\mu =  {\bar{d}\gamma_\mu\gamma_5u}$ is the (electrically charged) axial-vector current,  and $p_\mu$  is the four-momentum carried by the current. 
In the hadronic sector, the axial current can be expressed in terms of the pion field as $A_\mu = -\sqrt{2}f_\pi\partial_\mu\pi^+$.
In this work we consider two sum rules for the form factor $\Pi_0$, i.e the FESR considering two kernel values: 
one sum rule with a kernel $K=1$ and other with kernel $K=s$.%
    \footnote{alternatively we can use one sum rule for $\Pi_0$ and one for $\Pi_1$ with the same kernel $K=1$.} 
Therefore we need the OPE series up to order $s^{-2}$,
\begin{align}
 \left.\Pi_0(s)\right|_{\text{QCD}} =&
 -\frac{1}{4\pi^2}\ln(-s/\mu^2)
           -  2m_{q}\frac{\langle \bar{q} q \rangle}{s^2}
            +\frac{1}{12\pi} \frac{\langle \alpha_s \,G^2 \rangle}{s^2}, 
\end{align}
where $m_{q} \equiv \frac{1}{2}(m_u + m_d)$ is the average light quark mass, $G^2 \equiv G^{a\,\mu\nu}  G^a_{\mu\nu}$ and  $\langle \bar{q} q \rangle \equiv \frac{1}{2}(\langle\bar{u} u \rangle  + \langle \bar{d} d\rangle)$ the average light quark condensate.
The hadronic sector gives $\left.\Pi_0(s)\right|_{\text{Had}} = -2f_\pi^2/(s-m_\pi^2)$.

The second current correlator to be considered is $\Pi_{5 \nu} $,  involving an axial-vector current and its divergence
\begin{equation}
\Pi_{5 \nu} (p) \;=\; i \int d^{4} x \, e^{i p x} \,
\langle 0|T\,[ i\partial^\mu A_{\mu}(x) \,  A_{\nu}^{\dagger}(0)]|0 \rangle  
\;=\; p_\nu\, \Pi_5(p^2).
\label{Pi5}
\end{equation}
In this case, by the use of the equation of motion in QCD as well as in the hadronic sector, the divergence of the axial-vector current turn to be $\partial^\mu A_{\mu} =2m_q \,\bar{d}i\gamma_5 u$ in the QCD sector, and $\partial^\mu A_{\mu} = \sqrt{2}f_\pi^2m_\pi^2\pi^+$ in the hadronic sector. 
As expected, the divergence of the axial-vector current is proportional to the pseudoscalar current.
The expressions for $\Pi_5(s)$ in the hadronic sector and in QCD up to order $1/s$ is given by
\begin{align}
    \Pi_5(s)|_\text{\tiny QCD}=& - \frac{3}{2 \pi^2} \, {m_{q}}^2\, \ln(-s/\mu^2) +
    4m_q  \frac{\langle \bar{q} q \rangle}{s},
\end{align}
where we can see that the leading terms are proportional to the quark mass. 
The hadronic sector is given by $\left.\Pi_5(s)\right|_{\text{Had}} = -2f_\pi^2m_\pi^2/(s-m_\pi^2)$.

The third form factor we will consider is the one formed by the correlation of the axial-vector divergence
\begin{equation}\label{psi5}
    \psi_5(p^2) = i\int d^4x\,e^{ipx}\langle 0|T\,[\partial^\mu A_\mu(x)\,\partial^\nu A^\dagger_\nu(0) ]|0\rangle\,.
\end{equation}
Its QCD expression  to order $1/s$ is given by
\begin{equation}
    \psi_5(s)|_\text{\tiny QCD} = - \frac{3}{2 \pi^2} \, {m_{q}}^2\,s \ln(-s/\mu^2) 
 +4m_{q}^3\frac{\langle \bar{q} q \rangle}{s}
 -\frac{{m_{q}}^2}{2\pi} \frac{\langle\alpha_s\, G^2 \rangle}{s}
 ,
\end{equation}
and in the hadronic sector,  
$\psi_5(s)|_\text{\tiny QCD} = -f_\pi^2m_\pi^4/(s-m_\pi^2)$.\\

The three correlators are related through the Ward identities $p^\mu\Pi^A_{\mu\nu}(p)= \Pi_{5\nu}(p)+\langle\bar u\gamma_\nu u\rangle -\langle\bar d\gamma_\nu d\rangle$ and $p^\nu\Pi_{5\nu}(p)=\psi_5(p^2)+4m_q\langle\bar qq\rangle$.
The selected sum rules avoid the possibility of repeated information by choosing two sum rules for $\Pi_0$ with kernels $K=1$ and $K=s$, and one sum rule for $\Pi_5$ and $\Psi_5$ with a kernel $K=1$.
Notice that in this pseudoscalar channel we only consider the lowest state, which is the pion. 
If we do not include the next state, the $a_1$ resonance, the sum rules give a threshold $s_0$ smaller so the $a_1$ pole is located outside the contour of Fig. \ref{pacman}.  

Applying the sum rules from Eq. (\ref{FESR}) we obtain
\begin{align}
2 f_\pi^2 
    &= \frac{1}{4  \pi^2}s_0, \label{FESR01}\\
2 f_\pi^2 \, m_\pi^2 
    &= \frac{1}{8\pi^2}\,s_0^2 
        - 2m_{q} \,\langle \bar{q} q \rangle 
        - \frac{1}{12\pi}\langle \alpha_s G^2 \rangle ,  
        \label{FESR02}\\
2 f_\pi^2 m_\pi^2 
    &= - 4m_q\langle \bar{q} q \rangle 
        +\frac{3}{2\pi^2}m_q^2\, s_0 ,\label{FESR03}\\
2f_\pi^2 m_\pi^4
    &= \frac{3 }{4 \pi^2}m_q^2\, s_0^2 - 4m_{q}^2 \, \langle \bar{q} q \rangle \, +\, \frac{1}{2\pi}m_q^2\,\langle \alpha_s G^2 \rangle,
\label{FESR04}
\end{align}
where higher order quark-mass corrections  $( {m_{q}}^2/s_0)$ were neglected.
Notice that Eq.~(\ref{FESR03}) is the Gell-Man--Oakes--Renner (GMOR) relation \cite{GellMann:1968rz,Bordes:2010wy}, including a higher order quark-mass correction, i.e. $\cal{O}$$(m_q^2)$.
We use as an input the charged pion mass and the pion decay constant, in order to obtain, as a result, all the other parameters.

\section{FESR in an external magnetic field}

The presence of an external magnetic field  modifies  current correlators in several ways.
First of all, the magnetic field interacts with quarks and hadrons through the minimal coupling. 
We must replace then the derivatives with the corresponding covariant derivative.
Since the axial-vector current carries positive electric charge $e$ (the elementary proton charge),  its  derivative is replaced by the covariant derivative $ D^\mu A_\mu  = [\partial^\mu-ie{\cal A}^\mu]A_\mu$, where ${\cal A}$ denotes the vector potential of the external magnetic field.
Hence, the new definition of the correlators in Eq. (\ref{Pi5}) and (\ref{psi5}) changes by the replacement of 
$D_\mu $ instead of $\partial_\mu$.

Notice that the presence of the magnetic field breaks locality, and therefore $\Pi(x,y)\neq\Pi(x-y)$. However, we are free to choose any particular frame if it is the same for the hadronic and the QCD sector. 
The Ward identities discussed in the previous section are preserved, and their expressions are the same but
contracting the $\Pi^A_{\mu\nu}(p)$ and $\Pi_{5\mu}(p)$ correlators with $p^\mu +{\cal A}^\mu(p)$ instead of $p^\mu$.
We consider here a constant homogeneous electromagnetic external in the Schwinger-Fock symmetric gauge ${\cal A}^\mu(x) = -\frac{1}{2}F^{\mu\nu}x_\nu$, and therefore the electromagnetic vector potential in momentum space is then ${\cal A}^\mu(p) = \frac{ie}{2}F^{\mu\nu}\frac{\partial}{\partial p^\nu}$.

Now, if we consider an homogeneous external magnetic field along the $z$ axis, the electromagnetic field tensor  can then be written in the convenient form
$F_{\mu\nu}=B\epsilon_{\mu\nu}^\perp$, 
with the perpendicular anti-symmetric tensor defined as $\epsilon^\perp_{\mu\nu}= g_{\mu 1}g_{\nu 2}-g_{\mu 2}g_{\nu 1}$.
This term will appear in all tensor structures and lead to the separation of vectors into parallel and perpendicular projections.
The  metric is splitted into $g_{\mu\nu}=g_{\mu\nu}^\parallel +g_{\mu\nu}^\perp$. 
With this definition we have, for example $p^2=p_\parallel^2+p_\perp^2$, with  $p_\perp^2 = -\boldsymbol{p}_\perp^2$.
The magnetic field therefore introduces several modifications in the tensor structure of the current correlators. 
Basically it consists of combinations of $p_\mu$ $g_{\mu\nu}$ and $\epsilon_{\mu\nu}^\perp$ which produces a rich variety of new independent components, usually associated with new condensates.
For instance, for $\Pi_{\mu\nu}^{A}(p)$  the possible structures are $g^\parallel_{\mu\nu}$, $\epsilon_{\mu\nu}^{\perp}$, $g_{\mu\nu}^\perp$, and pair combinations of $p^\parallel_\mu$, $p_\mu^\perp$ and $\epsilon_{\mu\nu}^\perp p^\nu$.
Similarly, the possible structures for $\Pi_{5\nu}(p)$ are the  three mentioned vectors.

In the hadronic sector we use the axial-vector current field description from chiral perturbation theory ($\chi$PT) in terms of charged pion fields 
$    A_\mu= -\sqrt{2}(f^\parallel_\pi D^\parallel_\mu+f^\perp_\pi D^\perp_\mu)\pi^+$, where the parallel and perpendicular components of the pion decay constant are related with the transverse velocity as $v_\perp^2=f_\pi^\perp / f_\pi^\parallel$.
In this case the covariant divergence of the axial-vector current is $D^\mu A_\mu=  \sqrt{2}f^\parallel_\pi\, m_\pi^2\,\pi^+$.
This relation is obtained from the new equations of motion for the charged pion, $(D_\parallel^2+v_\perp^2D_\perp^2 +  m_\pi^2)\pi^+ = 0$.

We will consider the frame $p_\perp=0$, and the specific structures considered are the ones proportional to $p_\mu^\parallel p_\nu^\parallel$ from the axial-axial correlator, and $p_\mu^\parallel$ from the pseudoscalar-axial correlator.\\

The presence of a magnetic field is to be reflected in the charged particle propagators. 
These will be expressed in a power series involving the magnetic field \cite{Chyi:1999fc}.  
The quark  and the pion green functions become  $G_q(x,y)=e^{i e_q\phi_q(x,y)}S(x-y)$ and $G_\pi(x,y)=e^{i e_\pi\phi(x,y)}D(x-y)$, respectively,
where $e_{i}\phi$ is the Schwinger phase with $e_{i}$ the corresponding particle-charge and $\phi(x,y)=-\frac{1}{2}F_{\mu\nu}x^\mu y^\nu$  in the symmetric gauge. 

The local part of the Green functions in Fourier space are then expanded in powers of $B$.
The only terms needed for  magnetic corrections in QCD are the following:
\begin{align}
    S_q^{(0)}(k) = &~ i\frac{\slashed{k}+m_q}{k^2-m_q^2}\\
    S_q^{(1)}(k) = &~ -\gamma_1\gamma_2 (e_q B) \frac{(\slashed{k}_{\parallel} + m_q)}{(k^2 - m_q^2)^2}\\
    S_q^{(2)}(k) = &~  \frac{2 i\, (e_q B)^2}{(k^2 - m_q^2)^4}  \left[ k_\perp^2 (\slashed{k} + m_q) - \slashed{k}_\perp ( k_{\parallel}^2 - m_q^2) \right],
\end{align}
where the superscript $(n)$ refers to the power in the field $B^n$.
In the case of the pion,  the only contribution is that of the  propagator at zero magnetic field $D_\pi^{(0)}(p)=i/(p^2-m_\pi^2)$. 
This is because the next term is $D_\pi^{(1)} = 0$,  and the other terms do not survive in the FESR under consideration.\\

Once all the aforementioned ingredients are considered, we have to calculate the same diagrams needed in the vacuum case, extended with the magnetic Green functions and the covariant derivative.
The integrals involved are usually multi-loop diagrams (one-loop in our case) which can be handled by the introduction of Feynman parameters in the usual way. 
There will be infrared divergences from the magnetic contributions, which are safely controlled by the magnetic quark masses. 
Hence, it is necessary to keep finite quark masses to leading order in expansions in terms of $m_q^2/s_0$.
It is important to notice that even in the chiral limit there will be magnetic mass generation. 
Before integrating in the Feynman parameter it is more convenient first to integrate in the external momentum through the FESR contour.
The magnetic contribution to the contour integral in the complex squared-energy s-plane is given by
\begin{equation}
\oint_{\text{C}(s_0)} \frac{ds}{2\pi i} \, s^{N-1} \left[\int_0^1dx\,\frac{f(x)}{[s - {\cal M}^2]^n}\right]
=c_{N,n}\theta_{N,n} \int_0^1 dx f(x)({\cal M}^2)^{N-n}\theta(s_0-{\cal M}^2)
, \label{InNs_0}
\end{equation}
where $f$ is an arbitrary function of $x$, and ${\cal M}^2=m_q^2/x(1-x)$ and $c_{N,n}$ is a rational number.
The usual Heaviside function is denoted as $\theta(\xi)$, and here we define a discrete theta function $\theta_{N,n}=1$ for $N\geq n$ and $\theta_{N,n}=0$ for $N< n$.
The fact that this integral vanishes for $N< n$ is one of the most important features in this approach. 
Higher order contributions in powers of the magnetic field enters only if we consider a FESR kernel with higher powers in $s$.

Some of the magnetic Feynman diagrams involved are infrared divergent for massless quarks because $f(x)$ is singular.
To deal with it, notice that the contour integral in Eq.~(\ref{InNs_0}) is non vanishing only if ${\cal M}^2 < s_0$. 
This condition leads to the modification of the Feynman parameters integration limits
\begin{equation}
\int_0^1dx \,f(x)\, ({\cal M}^2)^{N-n}\theta\left(s_0 - {\cal M}^2\right)=
\int_{x_-}^{x_+}dx \,f(x)({\cal M}^2)^{N-n}
\label{Inn}
\end{equation}
where the limits $x_\pm =\frac{1}{2}\left(1\pm\sqrt{1+4m_q^2/s_0}\right)$ handle the IR divergences.
After the expansion in quark masses, there will appear logarithmic terms.  
These logarithms are strictly magnetically generated and vanish for $B=0$.

\section{Results}

The FESR involving magnetic field corrections are
\begin{align}
    2f^\parallel_{\pi}{}^2
            =&\; 
        \frac{s_0}{4\pi^2}  
\label{SR1-eB} \\
    2f^\parallel_{\pi}{}^2 m_\pi^2 
        =&\;
        \frac{1}{8\pi^2}\left\{s_0^2
        -\frac{2}{9}(eB)^2\left[10\ln(s_0/{m_{q}}^2)-27\right]\right\}
        +m_{q}\langle\bar qq\rangle
        -\frac{1}{12\pi}\langle\alpha_s G^2\rangle \\
   2f^\parallel_{\pi}{}^2 m_\pi^2
        =&\;
        - 4m_q\langle\bar qq\rangle 
        +\frac{3}{2\pi^2}{m_{q}^2} s_0
            \label{SR3-eB}\\
    2f^\parallel_{\pi}{}^2 m_\pi^4
  =&\;
    \frac{3m_q^2}{4\pi^2}\left\{s_0^2-\frac{20}{27} (eB)^2\left[
    \ln(s_0/{m_{q}}^2)-1
    \right]\right   \}
    -4m^3_{q}\langle\bar qq\rangle
    +\frac{m_q^2}{2\pi}\langle\alpha_sG^2\rangle
 \end{align}
where all the terms are functions of the magnetic field. 
The restriction ${m_{q}}^2\ll s_0$, remains valid for all  values of $eB$ under consideration. 
There are six parameters to be determined, i.e. 
$m_{q}$, $m_\pi$, $f_\pi$, $s_0$, $\langle \bar qq\rangle$, and $\langle\alpha_sG^2\rangle$. 
Since there are  only four independent FESR,  two inputs are required.
In the case of vacuum, the pion's mass and decay width are the more precise inputs. 
The case with finite magnetic field requires different inputs for solving the magnetic evolution of the parameters.
The most studied object in this case is the quark condensate, which in our case is obtained from 
Nambu--Jona-Lasinio (NJL) model results \cite{Coppola:2018vkw}, which agree with LQCD \cite{Bali:2012zg}, so we will use as an input the NJL chiral condensate at finite magnetic field normalized by its value at $B=0$. 

We need another ``magnetic'' input. 
During the realization of this work there wasn't much choice in the literature for the range of values of the magnetic field we are taking into account, so we consider three possible scenarios:
\begin{enumerate}
\item
The magnetic evolution of the charged pion mass provided by NJL calculations \cite{Coppola:2018vkw}. 
\item
$m_{q}/m_\pi^2=$constant. 
This scenario is natural from the Nambu-Goldstone realization of chiral $SU(2)\times SU(2)$ symmetry where we have $m_\pi^2 =2{\cal {B}} m_{q}$. 
So here we assume that ${\cal B}$ is independent of the magnetic field.
\item
$m_{q}=$ constant.
\end{enumerate}

\begin{figure}[b]
\begin{minipage}{.47\textwidth}
\includegraphics[scale=.59]{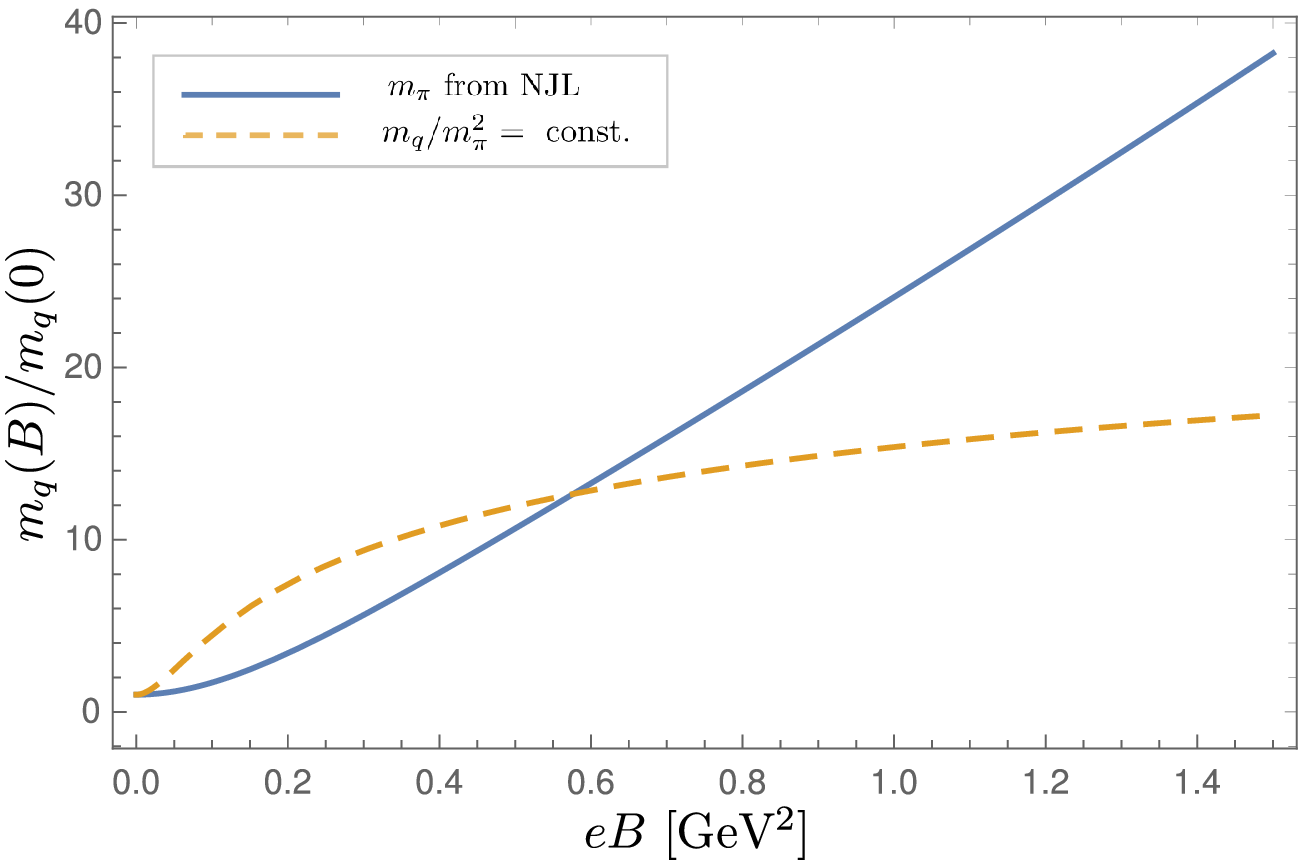}
\caption{ 
Magnetic evolution of the light quarks average mass. 
The plots show two possible inputs.}
\label{plotmq}
\end{minipage}
\hspace{.03\textwidth}
\begin{minipage}{.47\textwidth}
\includegraphics[scale=.58]{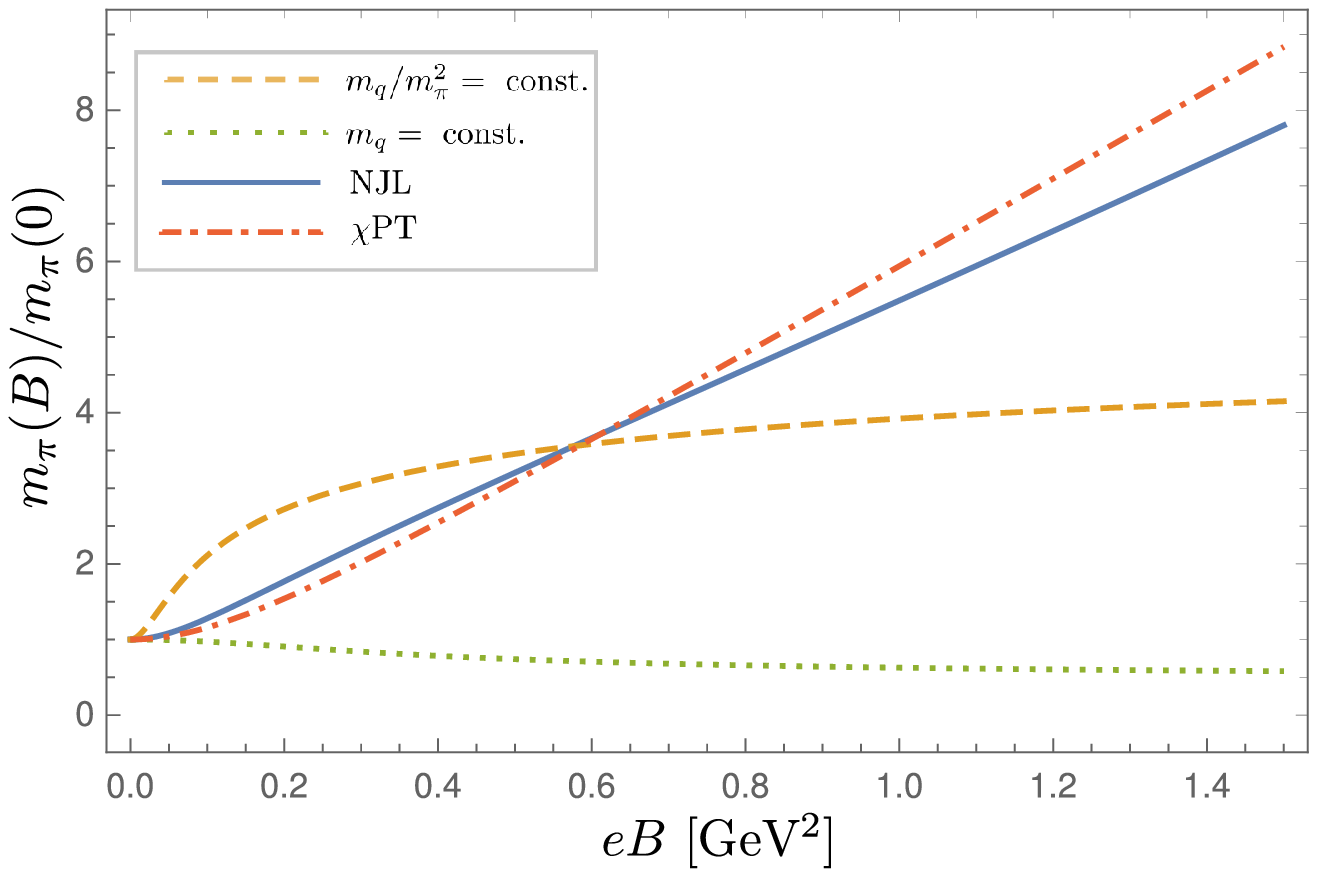}
\caption{ Magnetic evolution of the pion mass. 
The plot shows two possible inputs and results from NJL and $\chi$PT.}
\label{plotmpi}
\end{minipage}
\end{figure}

Figure~\ref{plotmq} shows the magnetic  quark mass  evolution and Fig.~\ref{plotmpi} shows the magnetic evolution of the pion mass. 
The similitude between the behaviour of both masses from the condition $m_q/m_\pi^2=$ constant (dashed  line) is expected.
The case with the pion mass obtained from NJL model or LQCD seems to be the same situation (continuous line), however both masses evolve differently and do not follow the relation $m_q \propto m_\pi^2$.
It is worth mentioning that the GMOR relation is broken here as stronger the magnetic strength.
This can be seen from Eq. (\ref{SR3-eB}),  where the presence of last term turn to be more relevant as the magnetic field increases.
In Fig.~\ref{plotmpi} we also show the result using $\chi$PT from \cite{Andersen:2012zc} to compare with.
All the pion mass results increase with the magnetic field, except for the case of constant $m_{q}$. 
This result reinforces the importance of the magnetic field behavior of $m_{q}$.

\begin{figure}
\begin{minipage}{.47\textwidth}
\includegraphics[scale=.59]{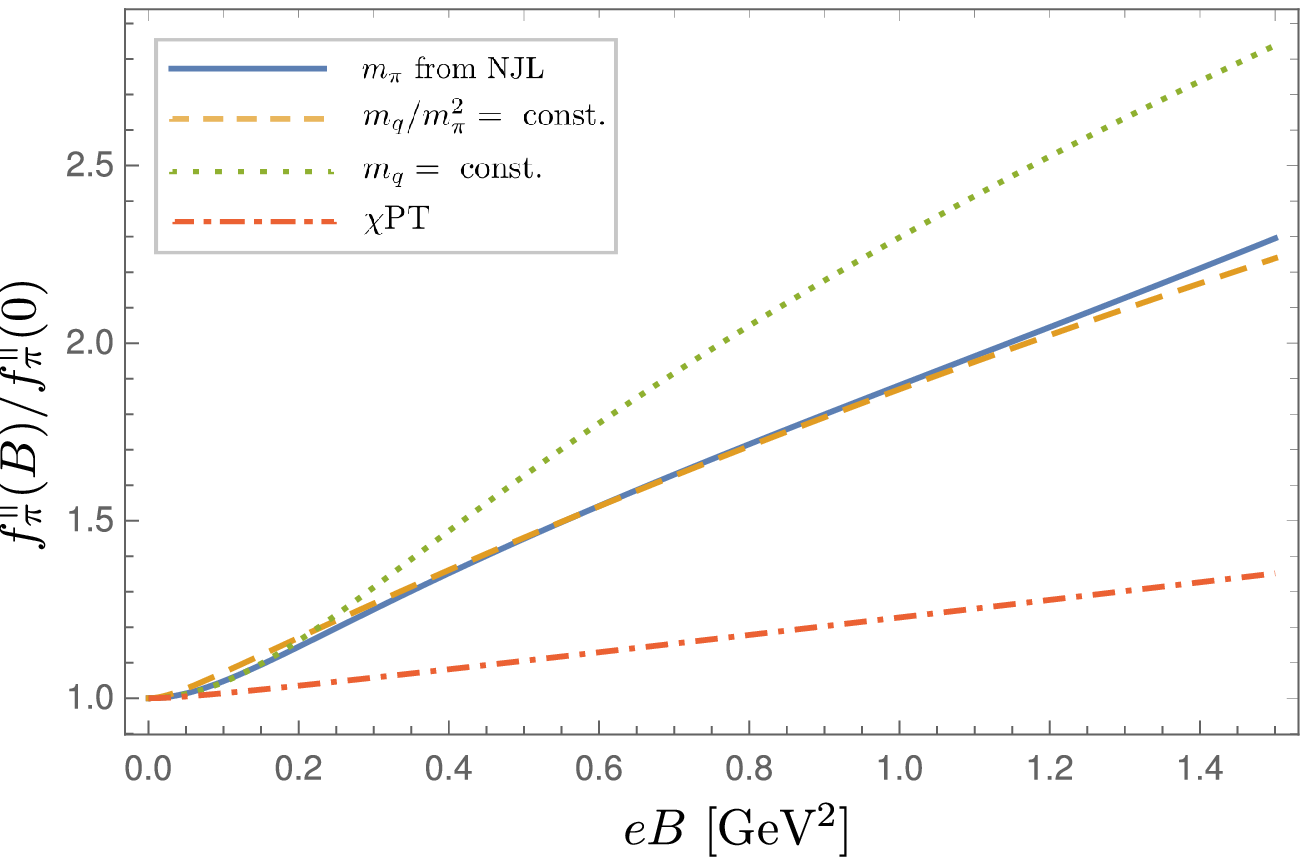}
\caption{Magnetic evolution of $f_\pi^\parallel$ (or equivalently $\sqrt{s_0}$), using three possible inputs and compared with $\chi$PT results.}
\label{plotfpi}
\end{minipage}
\hspace{.03\textwidth}
\begin{minipage}{.47\textwidth}
\includegraphics[scale=.57]{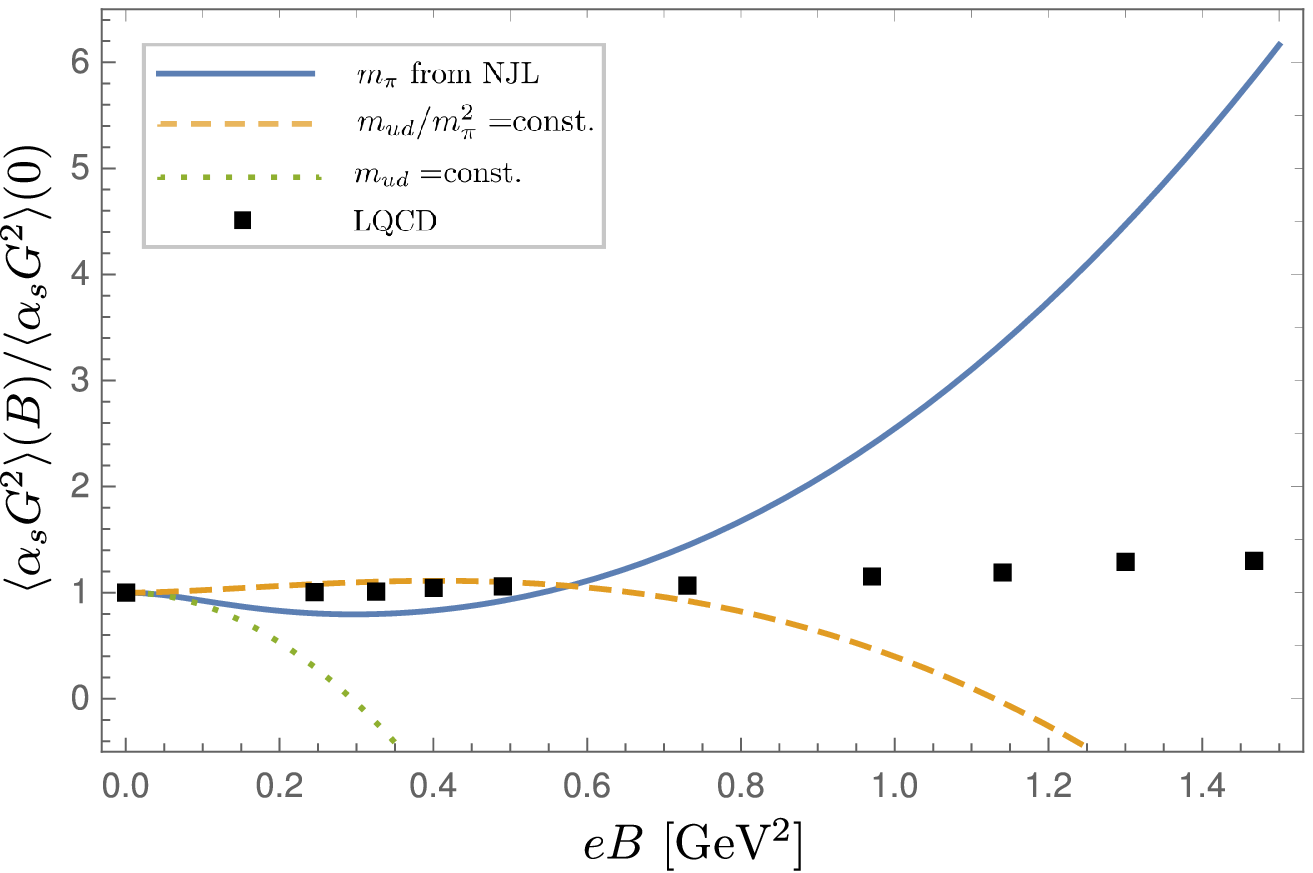}
\caption{Magnetic evolution of gluon condensate considering three possible inputs and compared with LQCD results.}
\label{plotGG}
\end{minipage}
\end{figure}

Figure \ref{plotfpi} shows the magnetic evolution of the  charged pion decay constant (parallel component) considering the three inputs described previously, compared with the result of the charged pion decay constant from $\chi$PT \cite{Andersen:2012zc}.
Notice that the two first conditions generates almost the same result for the decay constant. 
Also notice from Fig.~\ref{plotmpi} that the pion mass from $\chi$PT is very similar from the one obtained in NJL model, however the pion decay constant from $\chi$PT differs from the results obtained  in this work.
The case of constant quark mass presents a faster growing of the magnetic evolution of the decay constant.

Because of Eq. (\ref{SR1-eB}), the same plot describes the magnetic evolution of $\sqrt{s_0}$. 
This plot allows us to explore the validity of the magnetic field strength, because the $a_1$ resonance must be included if $\sqrt{s_0}\gtrsim 1.2$~GeV, . 
Considering that $\sqrt{s_0(0)}= 0.82$~GeV we can see that  $\sqrt{s_0(B)}\approx1.2$~GeV for $eB\approx 0.5$~GeV in the case of the results obtained using the conditions $m_\pi(B)$ from NJL, or $m_q/m_\pi^2 =$~constant.
In the case of the condition $m_q=$~constant, the maximum magnetic field strength $eB\approx 0.35$~GeV.
The increasing hadronic threshold confirms a previous work with FESR in the chiral limit \cite{Ayala:2015qwa}.

Figure~\ref{plotGG} shows  the magnetic evolution of the gluon condensate in the three conditions used as inputs and including lattice results to compare with.
For a constant quark mass the gluon condensate drops dramatically. This strongly suggests that a constant quark mass is not a valid approximation, although other works has founded also a decreasing behavior of the gluon condensate \cite{Agasian:1999sx}.
Although it seems to turn negative, as we pointed previously, the maximum magnetic field strength for the case with constant quark mass is $\sim 0.35$~GeV.

The other values seems to present completely different behaviors. 
The NJL pion mass as an input gives us an initially decreasing and then increasing gluon condensate, which is in accordance with the results in \cite{Ayala:2015qwa}.
With the condition of having a constant quotient $m_{q}/m_\pi^2$, the gluon condensate increases slightly for then decreasing gently with increasing magnetic field. 
Compared with Lattice results seems to deviate a for high magnetic field. 
However, the maximum allowed magnetic field value is $\sim 0.5$~GeV. 
The deviation in this region is less than 15\%, so it do not differ so much from lattice results.

The results obtained here and the detailed procedure are explained in \cite{Dominguez:2018njv}.  
There are many structures and new condensates that can be explored and were skipped in this work. 
One of the interesting structures is the separation of the parallel and perpendicular projections of the gluon condensate, separated in chromo-electric and chromo-magnetic components.
Another condensate we did not consider here is the polarization of the quark condensate $\langle \bar q\sigma_{12} q\rangle$.
We are working on these topics and will be reported elsewhere.

\section*{Aknowledgement}
The authors acknowledge 
financial support from Fondecyt (Chile) under grants 1190192, 1170107 and 1200483.
\section*{References}

\bibliographystyle{iopart-num}
\bibliography{bib}

\providecommand{\newblock}{}
\begin{thebibliography}{1}
\expandafter\ifx\csname url\endcsname\relax
  \def\url#1{{\tt #1}}\fi
\expandafter\ifx\csname urlprefix\endcsname\relax\def\urlprefix{URL }\fi
\providecommand{\eprint}[2][]{\url{#2}}

\bibitem{Dominguez:2018njv}
Dominguez C, Loewe M and Villavicencio C 2018 {\em Phys. Rev. D\/} {\bf 98}
  034015 (\textit{Preprint} \eprint{{arXiv:1806.10088}})

\bibitem{GellMann:1968rz}
Gell-Mann M, Oakes R and Renner B 1968 {\em Phys. Rev.\/} {\bf 175} 2195--2199

\bibitem{Bordes:2010wy}
Bordes J, Dominguez C, Moodley P, Penarrocha J and Schilcher K 2010 {\em
  JHEP\/} {\bf 05} 064 (\textit{Preprint} \eprint{arXiv:1003.3358})

\bibitem{Chyi:1999fc}
Chyi T~K, Hwang C~W, Kao W, Lin G~L, Ng K~W and Tseng J~J 2000 {\em Phys. Rev.
  D\/} {\bf 62} 105014 (\textit{Preprint} \eprint{arXiv:hep-th/9912134})

\bibitem{Coppola:2018vkw}
Coppola M, Gómez~Dumm D and Scoccola N 2018 {\em Phys. Lett. B\/} {\bf 782}
  155--161 (\textit{Preprint} \eprint{arXiv:1802.08041})

\bibitem{Bali:2012zg}
Bali G, Bruckmann F, Endrodi G, Fodor Z, Katz S and Schafer A 2012 {\em Phys.
  Rev. D\/} {\bf 86} 071502 (\textit{Preprint} \eprint{arXiv:1206.4205})

\bibitem{Andersen:2012zc}
Andersen J~O 2012 {\em JHEP\/} {\bf 10} 005 (\textit{Preprint}
  \eprint{arXiv:1205.6978})

\bibitem{Ayala:2015qwa}
Ayala A, Dominguez C, Hernandez L, Loewe M, Rojas J~C and Villavicencio C 2015
  {\em Phys. Rev. D\/} {\bf 92} 016006 (\textit{Preprint}
  \eprint{arXiv:1504.01308})

\bibitem{Agasian:1999sx}
Agasian N~O and Shushpanov I 2000 {\em Phys. Lett. B\/} {\bf 472} 143--149
  (\textit{Preprint} \eprint{hep-ph/9911254})

\end{thebibliography}

\end{document}